\documentclass[aps,prb,superscriptaddress]{revtex4-1} 

\usepackage{amsmath,amssymb}
\usepackage{graphicx}
\usepackage{epsfig}
\usepackage{color}
\usepackage[sort&compress]{natbib}

\begin{document}

\title{Polarization-resolved microscopy through scattering media \emph{via} wavefront shaping.}

\author{Hilton B. de Aguiar}
\email{h.aguiar@fresnel.fr}
\affiliation{Aix-Marseille Universit\'{e}, CNRS, Centrale Marseille, Institut Fresnel UMR 7249, 13013 Marseille, France}
\author{Sylvain Gigan}
\affiliation{Laboratoire Kastler Brossel, UMR 8552 of CNRS and Universit\'{e} Pierre et Marie Curie, 24 rue Lhomond, 75005 Paris, France}
\author{Sophie Brasselet}
\email{sophie.brasselet@fresnel.fr}
\affiliation{Aix-Marseille Universit\'{e}, CNRS, Centrale Marseille, Institut Fresnel UMR 7249, 13013 Marseille, France}

\begin{abstract}
Wavefront shaping  has revolutionized imaging deep in scattering media~\cite{Mosk2012,Horstmeyer2015}, being able to spatially~\cite{Vellekoop2007,Yaqoob2008,Popoff2010} and temporally~\cite{Katz2011,McCabe2011,Aulbach2011} refocus light through or inside the medium. However, wavefront shaping is not compatible yet with polarization-resolved microscopy given the need of polarizing optics to refocus light with a controlled polarization state. Here, we show that wavefront shaping is not only able to restore a focus, but it can also recover the injected polarization state without using any polarizing optics at the detection. This counter-intuitive effect occurs up to several transport mean free path thick samples, which exhibit a speckle with a completely scrambled state~\cite{Bicout1994,Xu2005}. Remarkably, an arbitrary rotation of the input polarization does not degrade the quality of the focus. This unsupervised re-polarization --- out of the originally scrambled polarization state --- paves the way for polarization-resolved structural microscopy~\cite{Brasselet2011} at unprecedented depths. We exploit this phenomenon and demonstrate second harmonic generation (SHG) structural imaging of collagen fibers in tendon tissues behind a scattering medium.\\
\end{abstract}

\maketitle 

\begin{figure}[ht]
    \centering
    \epsfig{file=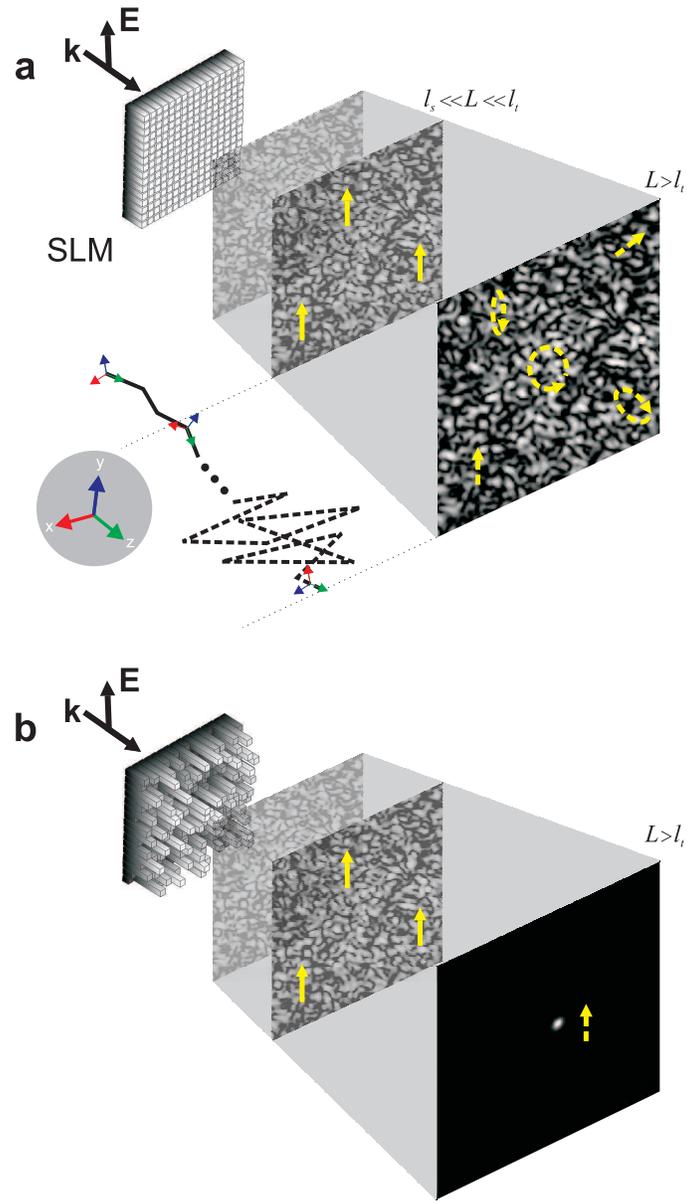,width=0.5\columnwidth}
    \caption{
		\textbf{Illustration of polarization state scrambling during propagation in a scattering medium and principle of polarization revival \emph{via} wavefront shaping}.
		When light propagates in a scattering medium, the wavefront is rapidly deformed into a speckle pattern, but polarization scrambling occurs on a different length scale. In the forward scattering regime (as typically found in biological tissues), when the thickness $L$ is much less the transport mean free path $l_t$, forward scattering events mainly conserves the initial polarization state. In the diffusive regime $L>l_t$, (a) for an unshaped wavefront, the polarization of the speckle gradually scrambles and lacks any resemblance with the input one. The bottom panel illustrates how during propagation each forward scattering conserves polarization (continuous line) and how polarization is mixed when entering the diffusive regime (dashed). 
		In contrast, we observe that (b) an optimal wavefront shaped by a spatial light modulator (SLM) is not only able to refocus light, but also recovers the original polarization state even without polarizing optics in the detection. 
}
    \label{fig1}
\end{figure}

Accessing the structural organization of molecular assemblies is an important aspect of biological imaging. For instance, this organization determines important functions such as the formation of bio-filaments, but also contributes to disorders such as in the formation of amyloids in neurodegenerative diseases. Polarization-resolved microscopies are able to provide molecular structural insights beyond the diffraction-limited scale~\cite{Brasselet2011}, and, being non-invasive and non-destructive, could in principle be used for deep \emph{in-vivo} structural imaging. Unfortunately, light propagation through multiply scattering media, like thick biological tissues, inherently scrambles polarization states. After injection of a well-defined polarization state and wavevector direction, each successive scattering event rotates the polarization~\cite{Xu2005}. Thus, after multiple scattering events the polarization is ultimately scrambled and the wavefront concomitantly distorted generating a speckle (schematically sketched in Fig. \ref{fig1}.a). This unavoidable scrambling makes molecular structural imaging impossible deep in biological media. Most importantly, it is highly detrimental for polarization-resolved studies because no predetermined relation with the incoming polarization state exists anymore: this is indeed a prerequisite for the analysis of the polarized response and its relation to structural modeling~\cite{Brasselet2011}.

Wavefront shaping has been introduced to overcome penetration depths limits in scattering media~\cite{Mosk2012}. By coherently controlling the speckle pattern generated from multiple scattering events, one can increase the energy density at targeted positions building a constructive interference locally. Wavefront shaping allows refocusing light through~\cite{Vellekoop2007}, or inside~\cite{Judkewitz2013,Chaigne2013,Horstmeyer2015}, scattering media thanks to the manipulation of the wavefront phase or amplitude~\cite{Akbulut2011,Conkey2012}. Nevertheless, the analysis of polarization effects on wavefront shaping experiments has been often disregarded because, in principle, there are no correlations among the $t^{ij}_{mn}$ elements of the vectorial transmission matrix $\mathbf{t}$~\cite{Popoff2010,Tripathi2012}, relating input field $E^i_n$ with polarization state $i$ to output field $E^j_m$ with polarization state $j$. Thus, in order to control the output polarization state, one needs to perform polarization sensitive measurements of the transmission matrix~\cite{Guan2012,Park2012}.

Here, we demonstrate an effect which enables polarization-resolved imaging through scattering media, and forms the basis for prospective deep structural molecular imaging: polarization revival out of a polarization-state-scrambled speckle (Fig. \ref{fig1}.b). 

\begin{figure}[ht]
    \centering
    \epsfig{file=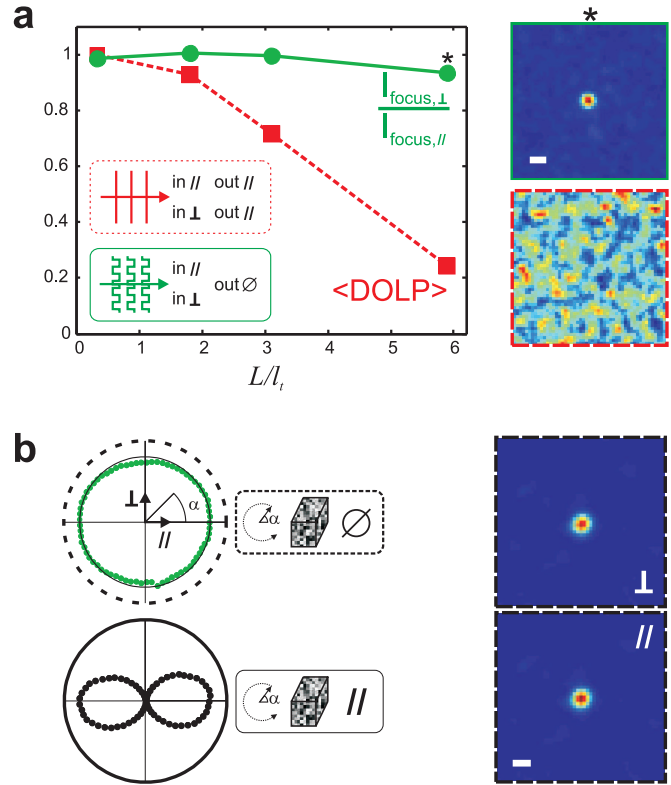,width=0.5\columnwidth}
		\caption{
		\textbf{Experimental quantification of polarization revival}.
			(a) Left panel: averaged degree of linear polarization of the speckle ($\left\langle \text{DOLP}\right\rangle$, red squares) and refocus polarization ratio after wavefront shaping ($\frac{\text{I}_\perp}{\text{I}_\parallel}$, corresponding to the same wavefront but rotated input polarization, green circles), as a function of optical depths $L/l_t$. Polarization scrambling of the speckle occurs for length of the order of $l_t$. Nevertheless, at depths of several $l_t$, after shaping, the refocus intensity survives a change of $90^\circ$ of the input polarization.
			(Right panels) speckle image after (top) and before (bottom) the wavefront shaping procedure. The scattering media are made of 5~$\mu$m-diameter polystyrene beads.
			(b) Similar experiments performed using opaque 1~mm-thick acute brain slice coronal cross section as scattering medium. (top left panel) Non-analyzed refocus intensity (green circles) upon rotation of the excitation field. (right panels) Images show the refocus at two input polarization state ($\perp \text{and} \parallel$) with the same intensity scale. (bottom left panel) The refocus polarization state purity is evaluated by placing an analyzer and observing an extinction (black circles).
			Scale bars: $1~\mu$m. 
		}
    \label{fig2}
\end{figure}

We first discuss aspects related to the speckle without wavefront shaping. We performed experiments at controlled optical thickness ($L/l_t$) using phantoms with scattering properties similar to biological media (see Methods). To aid the discussion, the inset of Fig. \ref{fig2}.a shows simplified schematics of the polarization state combinations used. Fig. \ref{fig2}.a (red dashed) shows the spatially output-averaged Degree of Linear Polarization (DOLP=$\left(\text{I}_\parallel-\text{I}_\perp\right)/\left(\text{I}_\parallel+\text{I}_\perp\right)$, where I is intensity) of the speckle observed outside the medium. As $L/l_t$ increases, the speckle polarization state decreases its correlation with the initial linearly polarized one as evidenced by the decrease in $\left\langle \text{DOLP}\right\rangle$. 

Conversely, when performing wavefront shaping the outcome is very different. To find the optimal wavefronts, a \emph{broadband} transmission matrix is first acquired~\cite{Popoff2010} (see Methods) \emph{without any selection of output polarization state} using the experimental layout shown in Fig. \ref{fig4}.a (Fig. \ref{fig2}.a inset, continuous line). In practice, this is equivalent to acquire experimental transmission matrix elements $t^{(exp)}_{mn}=t^{xx}_{mn}+t^{yx}_{mn}$. Once the transmission matrix is known, we can selectively refocus light at targeted positions (Fig. \ref{fig2}.a, upper right panel). Surprisingly, upon rotation of the input polarization state the refocus does not degrade. We further quantified this resilience to a change in input state as intensity ratio between the two injected polarization states ($\text{I}_\perp /\text{I}_\parallel$) shown in Fig. \ref{fig2}.a (continuous line). Clearly, only minute degradation of the refocus intensity ratio is seen reaching depths in the diffusive regime ($L/l_t>1$), in striking contrast with the DOLP measurements. A similar effect was observed through an opaque 1~mm-thick brain slice, where the dependence of the focus intensity with respect to the incident polarization is almost constant (Fig \ref{fig2}.b, top left panel).

Remarkably, the optimally shaped wavefront is able to refocus light with a well-defined polarization state, even though no output polarization state is privileged. The refocus is indeed highly polarized and correlated with the input one, as evidenced by the complete extinction of its intensity upon placement of an analyzer, as shown in Fig. \ref{fig2}.b (bottom left panel, see Supplementary Section 1 for polarization state evaluation of the refocus). 

\begin{figure}[ht]
    \centering
    \epsfig{file=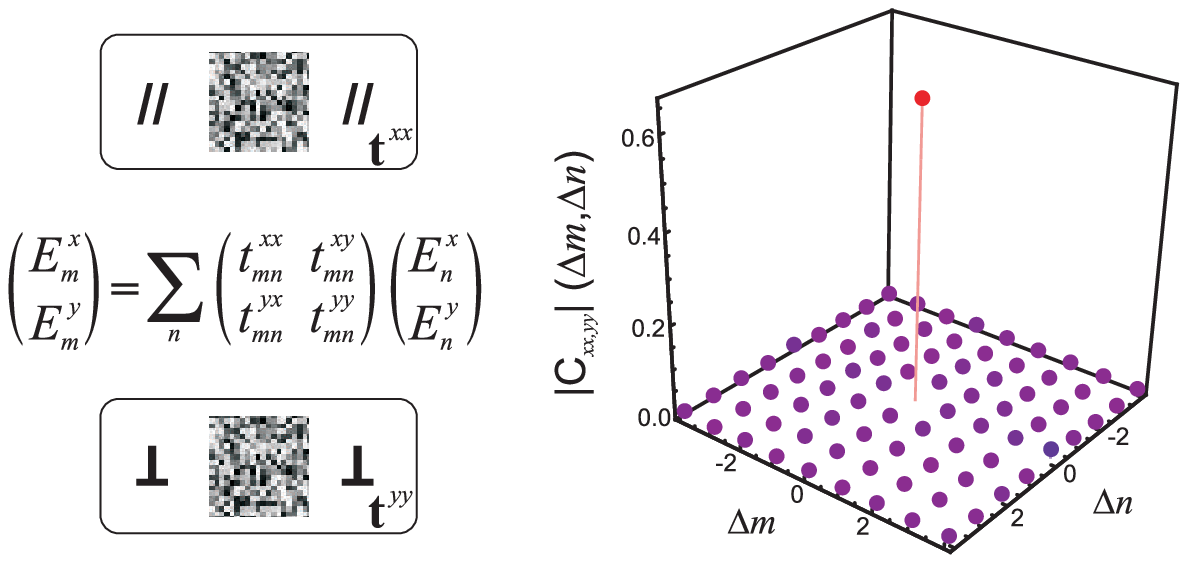,width=0.6\columnwidth}
		\caption{
		\textbf{Experimental quantification of vectorial transmission matrix correlations}.
		Cross-correlation of vectorial transmission matrix elements with polarization combinations $xx$ and $yy$ for $L/l_t\approx 6$. The peak confirms strong correlation between the matrix elements thus explaining the resilience of the focus to a polarization state change.
		}
    \label{fig3}
\end{figure}

The results above indirectly reveal correlations of the polarization-related part of the vectorial transmission matrix elements because they imply $t^{xx}_{mn}\approx t^{yy}_{mn}$. These correlations are formally shown in Fig. \ref{fig3} as cross-correlation of another set of measurements where $t^{xx}_{mn}$ and $t^{yy}_{mn}$ are independently measured for the same realization of disorder (with the polarization state combination used shown in Fig. \ref{fig3} left panels). In this analysis, which procedure is explained in Supplementary Section 2, the peak above the noise confirms that the diagonal elements of the vectorial transmission matrix are highly correlated. We conclude from these observations that $t^{(exp)}_{mn}\equiv t^{xx}_{mn}$, since the speckle from $t^{yx}_{mn}$ has a smaller contrast and is moreover weakly correlated with $t^{xx}_{mn}$ (see Supplementary Section 2 for correlation among elements). We further explored the reason for these effects and found out that the transmission matrix correlations and polarization state revival are related to the broadband nature of the acquired transmission matrix. Indeed, these effects are considerably decreased when the transmission matrix is taken with a monochromatic source, as expected~\cite{Guan2012,Park2012}~(see Supplementary Section 3 for similar experiments performed with monochromatic light). 

\begin{figure}[ht]
    \centering
    \epsfig{file=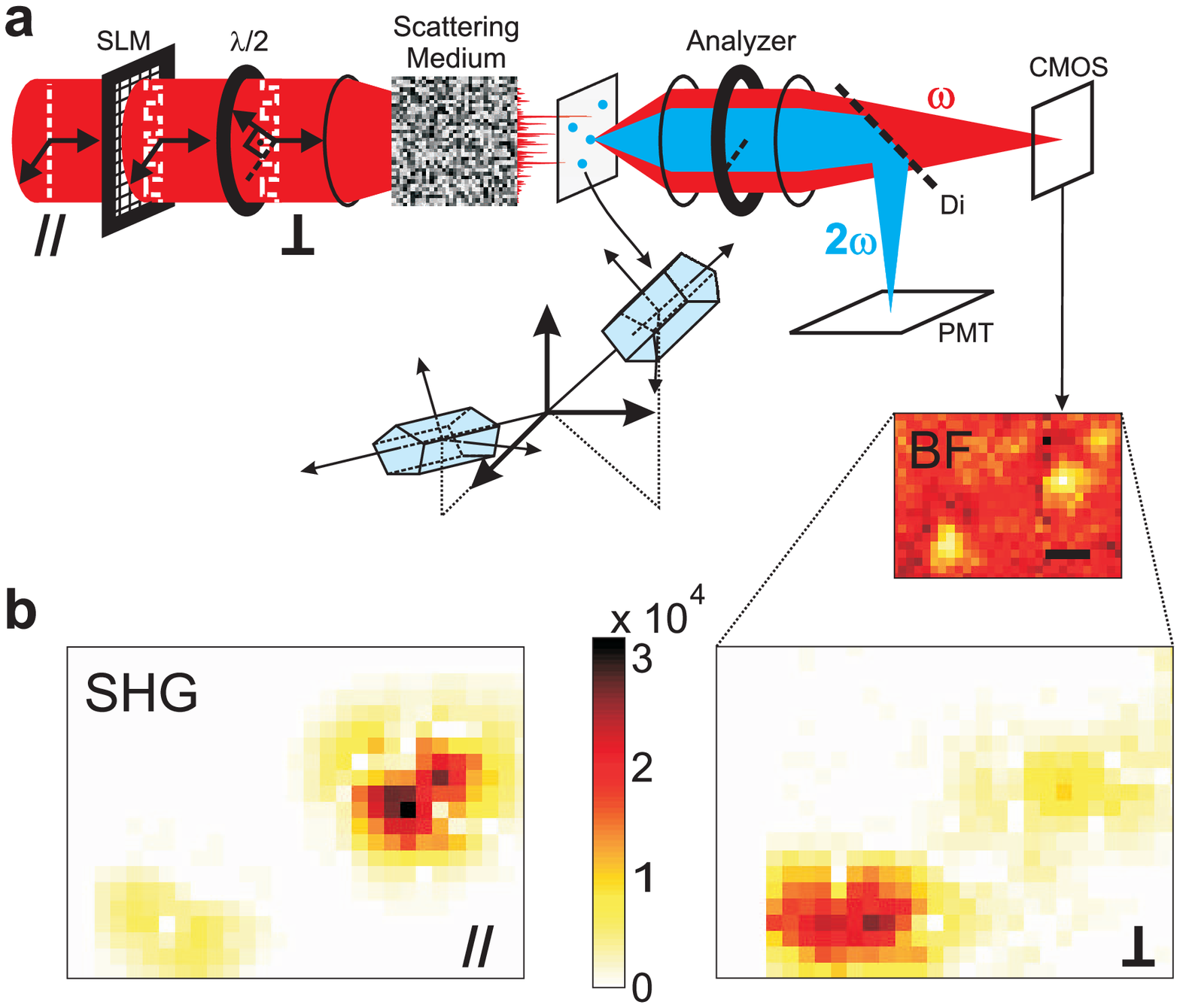,width=0.5\columnwidth}
    \caption{
		\textbf{Demonstration of structural imaging through scattering media exploiting transmission matrix correlations}.
		(a) Simplified experimental layout. Ultrashort pulse wavefronts are shaped by a SLM and focused on the scattering medium. The speckle transmitted by the scattering medium excites the nonlinear sources (nanoKTP) placed at a plane further imaged on a complementary metal-oxide semiconductor camera (CMOS) and a photomultiplier tube (PMT). 
		(b) Polarization-resolved SHG images. In a first step, the vectorial transmission matrix elements $t^{xx}_{mn}$ are acquired and used for raster scanning the refocus thus generating the SHG images (left panel). The inset in (a) shows the brightfield (BF) image at the same region of interest (ROI) where two particles can be seen. In a second step, only the excitation polarization is rotated and a second scan is taken (right panel, 5x rescaled) using the very same $t^{xx}_{mn}$ elements. 
		Scale bar: $1~\mu$m.
		}
    \label{fig4}
\end{figure}

We then exploited these correlations for demonstrating nonlinear structural imaging. Fig. \ref{fig4} shows the experimental layout and representative results of a wavefront shaping experiment. In a first step, a polarization combination is used where analyzer and incoming polarization state are aligned together, that is, a configuration measuring the $t^{xx}_{mn}$ elements. Note that the presence of the analyzer is not necessary for a wavefront shaping experiment, but it supports an unambiguous proof of structural imaging (see below). After acquiring the broadband transmission matrix (see Methods), we spatially refocus light in a raster scanning fashion and acquire in parallel the enhanced SHG signal from nanocrystals of potassium titanyl phosphate (nanoKTP). The dipolar character of nanoKTP ensures that no ambiguity exists in the polarization state evaluation. The refocused light generates efficient SHG from the nanoscopic sources as seen by the two nanoparticles in Fig. \ref{fig4}.b (left panel). In a second step, a second imaging scan using the very same optimal wavefronts is taken (Fig. \ref{fig4}.b, right panel), however with the incoming polarization state rotated 90$^\circ$. The refocus after the scattering medium still persists and highlights the second particle revealing the orientation-dependent nonlinear efficiency. This highlights the presence of nanoKTP crystals with different orientations, which are identified using the same wavefront. Any oriented nonlinear source could be probed in the same way, highlighting its molecular order organization in a scattering medium. A similar situation is found, for example, in coherent Raman imaging of biological membranes~\cite{Potma2003} or myelin sheath in neurons~\cite{Wang2005b}.

\begin{figure}[ht]
    \centering
    \epsfig{file=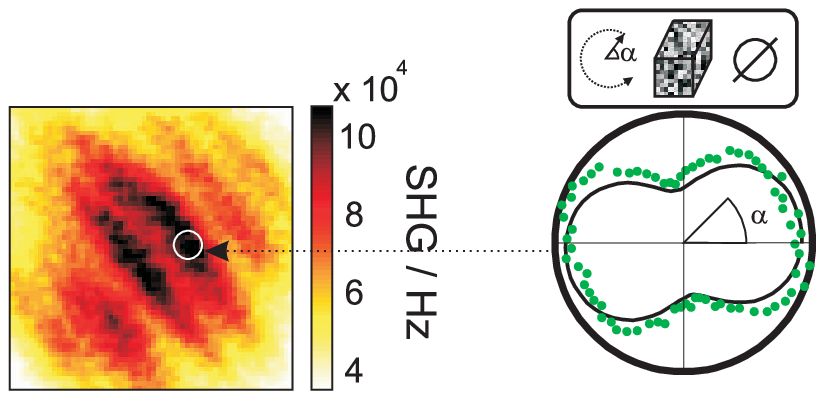,width=0.5\columnwidth}
    \caption{
		\textbf{Applications of transmission matrix correlations for biological specimens nonlinear structural imaging.}
		(left panel) Raster scanning using the memory effect generates SHG images from rat tail collagen tendon placed after a thin diffuser with a known transmission matrix. (right panel) By refocusing on a specific fiber, the intensity response of the SHG signal is recorded as the input angle varies. The continuous line (black) is a fit to the data (green circles) from which we retrieve the nonlinear susceptibilities values of the collagen fiber. The retrieved nonlinear susceptibilities reveals the molecular order of the fibers and are in excellent agreement with previous observations (see Supplementary Section 4).
		}
    \label{fig5}
\end{figure}

The implication of these observations for structural nonlinear imaging in biologically relevant specimens are shown in Fig. \ref{fig5}. In order to mimic a nonlinear microscopy experiment, we showcase SHG imaging of collagen fibers placed after the scattering medium with a characterized transmission matrix. Different from the previous experiment, here we exploit the memory effect to raster scan the refocus~\cite{Schott2015,Judkewitz2015}. This methodology is able to provide highly contrasted SHG images of rat tendon specimens where clear fibers can be seen running along the diagonal of the images (Fig. \ref{fig5}, left panel). In this demonstration, we do not use an analyzer to detect the SHG: in an eventual refocusing inside a scattering medium the nonlinearly generated photons polarization state would be scrambled~\cite{deAguiar2015}. Therefore, structural imaging is only insightful if performed in a non-analyzed configuration. By fixing the refocus position in a fiber and rotating the excitation polarization state, we can still perform a polarization-resolved study~\cite{Duboisset2012,Bioud2014}. Fig. \ref{fig5} (right panel) shows the measured SHG intensity (green circles) and a fit (black continuous line). The retrieved nonlinear optical constants of collagen are in excellent agreement with previously reported values~\cite{Stoller2002} (see Supplementary Section 4 for further information on the SHG analysis).

Finally, we expect these correlations to have profound implications for deep structural imaging in various approaches. For instance, acoustic-based methods~\cite{Wang2012,Judkewitz2013,Chaigne2013} are able to refocus in the diffuse regime ($L>~l_t$) with cellular resolution. This means that the acoustic signal from the obtained refocus could be able to perform structural analysis: If the absorbers are organized in an ordered way, a rotation of the refocus optical polarization, using the very same wavefront, should lead to a change in the acoustic signal.

\section*{METHODS}
\textbf{Optical set-up}. Ultrashort pulses (130~fs, 800~nm, 76~MHz repetition rate, Mira, Coherent) are steered onto a $256\times 256$ pixels reflective SLM (Boulder Nonlinear Systems). The SLM is imaged on the back focal plane of the focusing lens (achromatic lens, f=19~mm, AC127-019-B-ML, Thorlabs) with the scattering medium placed after it. The nanoKTP crystals are imaged by an objective (40X, 0.75~NA, Nikon) on a 12-bit CMOS camera (Flea3, Point Grey) and on a large area photon counting PMT (MP 953, PerkinElmer). The SHG signal is spectrally separated with suitable dichroic mirror (560~nm longpass, AHF Analysentechnik), shortpass (700~nm, FESH0700, Thorlabs) and bandpass (400~$\pm$~10~nm, Chroma Technology) filters. Additionally, a longpass filter (650~nm, FELH0650, Thorlabs) and a zero-order half-wave plate (46-555, Edmund Optics) --- placed in a motorized rotation stage --- are used for the excitation beam before the focusing lens. An analyser (WP25M-UB, Thorlabs) is used to select the detected polarization state whenever stated.

\textbf{Broadband transmission matrix measurement, wavefront shaping and nonlinear imaging methodology}. A thorough description of the methods used for acquisition of the transmission matrix can be found elsewhere~\cite{Popoff2010,Chaigne2013}. It consists of a two-step process: first the broadband transmission matrix of the system is acquired using ultrashort pulses. Once acquired, the transmission matrix elements are used for selective refocusing in a ROI. Briefly, for each input-output channel, phase and amplitude of the corresponding transmission matrix element are obtained by phase-shifting interferometry. The wavefront phase of a Hadamard base (input channel) is shifted in respect to a reference field in the range [$0-2\pi$] with its intensity recorded by a CMOS pixel (output channel). A Fourier transform is applied on the output channel interferogram of the $n$th Hadamard basis thus retrieving its phase and amplitude. Once all the Hadamard basis are measured, an unitary transformation is applied to obtain the transmission matrix in the canonical basis~\cite{Popoff2011a}. For acquiring the nonlinear images two methodologies were used. In Fig. \ref{fig4}, we raster scan the refocus at each spatial position in the ROI containing the nanoKTP and, in parallel, collecting the SHG signal integrated within the imaged plane with the large area PMT. In Fig. \ref{fig5}, we raster scan the refocus using the memory effect: a linear phase ramp is added to the wavefront, thus displacing the refocus in the image plane (with the SHG signal acquired in parallel). Number of controlled SLM segments for the experiments: $2^{12}$ (Fig. \ref{fig4}) and $2^{10}$ (Figs. \ref{fig2}, \ref{fig3}, \ref{fig5})

\textbf{Sample preparation}. Various types of scattering media were used: dispersions of $5~\mu$m-diameter polystyrene beads (Fig. \ref{fig2}.a and  \ref{fig3}), 1-mm thick brain slice (Fig. \ref{fig2}.b) --- both fixed in agarose solution --- 1~mm-thick parafilm film~\cite{Hsieh2010} (Fig. \ref{fig4}) and a commercial diffuser (Fig. \ref{fig5}, $10^\circ$ Light Shaping Diffuser, Newport). The rat tendon (Fig. \ref{fig5}) was placed between two coverlips separated by a 120~$\mu$m-thick spacer and filled with agarose solution. The 1~mm-thick mouse brain coronal cross-section (Fig. \ref{fig2}.b) followed the same protocol, except the spacer was 1~mm-thick. The nanoKTP crystals (150~nm diameter) were previously characterized by various methods~\cite{Mayer2013} and were drop cast on a coverslip (170~$\mu$m-thickness). The scattering mean free path, $l_s$, of the scattering media in Fig. \ref{fig2}.a and \ref{fig3} ($l_s=53-69~\mu$m)  were determined by measuring the extinction of the laser through a thin dispersion slab of known thickness, with scattered light and ballistic light spatially isolated in the Fourier space~\cite{deAguiar2015}. Anisotropy values $g$ for calculating $l_t=l_s/(1-g)$ were obtained using the scattering pattern from Mie theory.

\section*{Acknowledgements} The authors thank Esben Andressen and Herve Rigneault for fruitful discussions and support, Arnaud Malvache and Rosa Cossart for providing the brain slices, Thierry Gacoin and Ludovic Meyer for providing the nanoKTP particles. This work has been supported by the FEMTO Network and contracts ANR-10-INBS-04-01 (France-BioImaging infrastructure network), ANR-11-INSB-0006 (France Life Imaging infrastructure network), and the A*MIDEX project ANR-11-IDEX-0001-02. S.G. is  funded by the European Research Council (grant 278025 - COMEDIA).

\section*{Author contributions} H.B.A. and S.B. conceived and developed the ideas with fundamental input of S.G. H.B.A. developed and performed all experiments, processed and analyzed the data. All authors contributed to the writing and editing of the manuscript.


\begin{thebibliography}{30}

\bibitem{Mosk2012} A. P. Mosk, A. Lagendijk, G. Lerosey, and M. Fink. Controlling waves in space and time for imaging and focusing in complex media. \emph{Nature Photon.}, \textbf{6}, 283--292, 2012.

\bibitem{Horstmeyer2015} R.~Horstmeyer, H.~Ruan, and C.~Yang. Guidestar-assisted wavefront-shaping methods for focusing light into biological tissue. \emph{Nature Photon.}, \textbf{9}, 563--571, 2015.

\bibitem{Vellekoop2007} I.~M.~Vellekoop and A.~P. Mosk. Focusing coherent light through opaque strongly scattering media. \emph{Opt. Lett.}, \textbf{32}, 2309--2311, 2007.

\bibitem{Yaqoob2008} Z. Yaqoob, D. Psaltis, M.~S. Feld, and C. Yang. Optical phase conjugation for turbidity suppression in biological samples. \emph{Nature Photon.}, \textbf{2}, 110--115, 2008.

\bibitem{Popoff2010} S.~M. Popoff, G.~Lerosey, R.~Carminati, M.~Fink, A.~C. Boccara, and S.~Gigan. Measuring the transmission matrix in optics: An approach to the study and control of light propagation in disordered media. \emph{Phys. Rev. Lett.}, \textbf{104}, 100601, 2010.

\bibitem{Katz2011} O. Katz, E. Small, Y. Bromberg, and Y. Silberberg. Focusing and compression of ultrashort pulses through scattering media. \emph{Nature Photon.}, \textbf{5}, 372--377, 2011.

\bibitem{McCabe2011} D.~J. McCabe, A. Tajalli, D.~R. Austin, P. Bondareff, I.~A Walmsley, S. Gigan, and B. Chatel. Spatio-temporal focusing of an ultrafast pulse through a multiply scattering medium. \emph{Nature Commun.}, \textbf{2}, 447, 2011.

\bibitem{Aulbach2011} J. Aulbach, B. Gjonaj, P.~M. Johnson, A.~P. Mosk, and A.~Lagendijk. Control of light transmission through opaque scattering media in space and time. \emph{Phys. Rev. Lett.}, \textbf{106}, 103901, 2011.

\bibitem{Bicout1994} D.~Bicout, C.~Brosseau, A.~S. Martinez, and J.~M. Schmitt. Depolarization of multiply scattered waves by spherical diffusers: Influence of the size parameter. \emph{Phys. Rev. E}, \textbf{49}, 1767--1770, 1994.

\bibitem{Xu2005} M.~Xu and R.~R. Alfano. Random walk of polarized light in turbid media. \emph{Phys. Rev. Lett.}, \textbf{95}, 213901, 2005.

\bibitem{Brasselet2011} S. Brasselet. Polarization-resolved nonlinear microscopy: application to structural molecular and biological imaging. \emph{Adv. Opt. Photon.}, \textbf{3}, 205--271, 2011.

\bibitem{Judkewitz2013} B. Judkewitz, Y.~Min Wang, R. Horstmeyer, A. Mathy, and C. Yang. Speckle-scale focusing in the diffusive regime with time reversal of variance-encoded light (trove). \emph{Nature Photon.}, \textbf{7}, 300--305, 2013.

\bibitem{Chaigne2013} T.~Chaigne, O.~Katz, A.~C~Boccara, M.~Fink, E.~Bossy, and S.~Gigan. Controlling light in scattering media non-invasively using the photoacoustic transmission matrix. \emph{Nature Photon.}, \textbf{8}, 58--64, 2013.

\bibitem{Akbulut2011} D.~Akbulut, T.J. Huisman, E.G. van Putten, W.L. Vos, and A.~P. Mosk. Focusing light through random photonic media by binary amplitude modulation. \emph{Opt. Express}, \textbf{19}, 4017--4029, 2011.

\bibitem{Conkey2012} D.~B. Conkey, A.~M. Caravaca-Aguirre, and R. Piestun. High-speed scattering medium characterization with application to focusing light through turbid media. \emph{Opt. Express}, \textbf{20}, 1733--1740, 2012.

\bibitem{Tripathi2012} S. Tripathi, R. Paxman, T. Bifano, and K.~C. Toussaint. Vector transmission matrix for the polarization behavior of light propagation in highly scattering media. \emph{Opt. Express}, \textbf{20}, 16067--16076, 2012.

\bibitem{Guan2012} Y. Guan, O. Katz, E. Small, J. Zhou, and Y. Silberberg. Polarization control of multiply scattered light through random media by wavefront shaping. \emph{Opt. Lett.}, \textbf{37}, 4663--4665, 2012.

\bibitem{Park2012} J.-H. Park, C. Park, H. Yu, Y.-H. Cho, and Y.-K. Park. Dynamic active wave plate using random nanoparticles. \emph{Opt. Express}, \textbf{20}, 17010--17016, 2012.

\bibitem{Potma2003} E.~O. Potma and X.~S. Xie. Detection of single lipid bilayers with coherent anti-Stokes Raman scattering (CARS) microscopy. \emph{J. Raman Spectrosc.}, \textbf{34}, 642--650, 2003.

\bibitem{Wang2005b} H. Wang, Y. Fu, P. Zickmund, R. Shi, and J.-X. Cheng. Coherent anti-Stokes Raman scattering imaging of axonal myelin in live spinal tissues. \emph{Biophys. J.}, \textbf{89}, 581--591, 2005.

\bibitem{Schott2015} S. Schott, J. Bertolotti, J.-F. L´eger, L. Bourdieu, and S. Gigan. Characterization of the angular memory effect of scattered light in biological tissues. \emph{Opt. Express}, \textbf{23}, 13505--13516, 2015.

\bibitem{Judkewitz2015} B.~Judkewitz, R.~Horstmeyer, I.~M. Vellekoop, I.~N. Papadopoulos, and C.~Yang. Translation correlations in anisotropically scattering media. \emph{Nature Phys.}, \textbf{11}, 684--689, 2015.

\bibitem{deAguiar2015} H.~B. de~Aguiar, P. Gasecka, and S. Brasselet. Quantitative analysis of light scattering in polarization-resolved nonlinear microscopy. \emph{Opt. Express}, \textbf{23}, 8960--8973, 2015.

\bibitem{Duboisset2012} J. Duboisset, D. A\"it-Belkacem, M. Roche, H. Rigneault, and S. Brasselet. Generic model of the molecular orientational distribution probed by polarization-resolved second-harmonic generation. \emph{Phys. Rev. A}, \textbf{85}, 043829, 2012.

\bibitem{Bioud2014} F.-Z. Bioud, P. Gasecka, P. Ferrand, H. Rigneault, J. Duboisset, and S. Brasselet. Structure of molecular packing probed by polarization-resolved nonlinear four-wave mixing and coherent anti-Stokes Raman-scattering microscopy. \emph{Phys. Rev. A}, \textbf{89}, 013836, 2014.

\bibitem{Stoller2002} P. Stoller, K.~M. Reiser, P.~M. Celliers, and A.~M. Rubenchik. Polarization-Modulated Second Harmonic Generation in Collagen. \emph{Biophys. J.}, \textbf{82}, 3330--3342, 2002.

\bibitem{Wang2012} Y. M. Wang, B. Judkewitz, C.~A. DiMarzio, and C. Yang. Deep-tissue focal fluorescence imaging with digitally time-reversed ultrasound-encoded light. \emph{Nature Commun.}, \textbf{3}, 928, 2012.

\bibitem{Popoff2011a} S.~M. Popoff, G.~Lerosey, M.~Fink, A.~C. Boccara, and S.~Gigan. Controlling light through optical disordered media: transmission matrix approach. \emph{New J. Phys.}, \textbf{13}, 123021, 2011.

\bibitem{Hsieh2010} C.-L. Hsieh, Y. Pu, R. Grange, G. Laporte, and D. Psaltis. Imaging through turbid layers by scanning the phase conjugated second harmonic radiation from a nanoparticle. \emph{Opt. Express}, \textbf{18}, 20723--20731, 2010.

\bibitem{Mayer2013} L. Mayer, A. Slablab, G. Dantelle, V. Jacques, A.-M. Lepagnol-Bestel, S. Perruchas, P. Spinicelli, A. Thomas, D. Chauvat, M. Simonneau, T. Gacoin, and J.-F. Roch. Single KTP nanocrystals as second-harmonic generation biolabels in cortical neurons. \emph{Nanoscale}, \textbf{5}, 8466--8471, 2013.

\end{thebibliography}
\end{document}